\documentclass{mn2e}
\usepackage{epsfig}
\title[\it{VI} photometry of AA~Dor]
  {New $VI_{C}$ photometry of the sdOB
   binary AA~Dor and an improved photometric model}
\author[R.W.Hilditch \it{et al.}]
  {R.W.~Hilditch,$^1$ D.~Kilkenny,$^2$ A.E.~Lynas-Gray$^3$ and G.~Hill$^4$\\
  $^1$School of Physics and Astronomy, University of St Andrews, St Andrews,
Fife KY16 9SS\\  
  $^2$South African Astronomical Observatory, Observatory, Cape, 
South Africa\\
  $^3$Department of Physics, University of Oxford, Oxford,
OX1 3RH\\
  $^4$18A Stratford St., Auckland 1, New Zealand\\} 
\begin{document}
\maketitle

\begin{abstract}
  New $VI_{C}$ CCD photometry, obtained with integration times of 20s,
of the sdOB+degenerate-dwarf eclipsing binary system AA~Dor has provided new
complete light curves with an \textit{rms} scatter about a mean curve of
$\pm\,0.004$\,mag.  These data are analysed with an improved
\textsc{Light2} light curve synthesis code to yield more accurate
determinations of the radii of both stars, the orbital inclination,
and the flux ratio between the two components. These radii are 
only a little different from the values derived 25 years ago from less
complete data, but the uncertainties on these values are improved by a
factor of two.  The apparent discrepancy remains between the surface
gravity of the sdOB primary star obtained from the light-curve solution
with the published spectroscopic orbit and that obtained from NLTE
analysis of high-resolution spectra of the sdOB star.

 The substantial reflection effect in the system is adequately
represented by the \textsc{Light2} code with a bolometric albedo of
unity in light curves extending from $0.35\,\mu\,m$ to $2.2\,\mu\,m$.
However, there are differences at individual wavelengths in the derived
albedo, which may indicate redistribution of flux from shorter wavelengths
into the $V$ and $I_{C}$ passbands.
\end{abstract}


\section{Introduction}

The blue star LB 3459 (HD269696; $V\simeq\,11.1\,m$,
$RA=05^{h}\,31^{m}\,40^{s}.3,\,
Dec=-69^{\circ}\,53^{'}\,02^{''}.2\,(2000)$), and situated at high
galactic latitude ($b=-32^{\circ}.18$), was discovered to be variable
in brightness by Kilkenny, Hilditch \& Penfold (1978).
Observations with the University of Cape Town high-speed photometer
revealed the light curve of an eclipsing binary (designated AA
Doradus) with an orbital period of 0.261 day and a primary eclipse of
depth 0.4 mag in the V-band. Spectroscopic and further photometric
observations followed over many years, and the present understanding
of the properties of this remarkable binary system is summarized in
Hilditch, Harries \& Hill (1996) (hereafter HHH), and in Rauch (2000)
where references to previous publications on this system may be found.

HHH obtained \'{e}chelle spectra of AA~Dor
from the Anglo- Australian Telescope and determined a velocity
semi-amplitude for the sdOB primary star of
$40.8\pm\,0.7\,km\,s^{-1}$. Many attempts to detect spectral
signatures of the secondary component from these data failed, and HHH
were forced to adopt a representative mass of $0.5\,M_{\odot}$ for the
sdO star in order to calculate the mass of the secondary star from the
derived mass function and the earlier light-curve analyses, and
establish the absolute sizes of the stars in this system. The sdOB star
seemed to be an unremarkable member of that class of evolved stars,
whilst the unseen secondary component at $0.09\,M_{\odot}$ seemed to
be a degenerate red dwarf of surface temperature $\simeq2000K$. The derived
surface gravity of the primary star was found to be
$log\,g=5.53\pm\,0.03$ which was in some disagreement with an earlier
NLTE analysis by Kudritski,\textit{et al.} (1982), where a value of
$5.3\pm\,0.2$ was derived. HHH recommended that a
new determination of $log\,g$ from spectral line profiles would help
to resolve this issue and permit a direct determination of the mass of
the sdOB star and therefore also the secondary component.

The more recent work by Rauch (2000) addressed this issue directly by
detailed NLTE analysis of high resolution optical and ultraviolet
spectra. Rauch obtained an effective temperature for the primary
component of $42000\pm\,1000\,K$, in agreement with Kudritski \textit{et al.},
and a value of $log\,g=5.21\pm\,0.10$ which clearly exacerbated the problem.
He also derived a mass for the sdO star of $0.330\pm\,0.006$ from 
comparison of the location of the primary in the $T_{eff}\,vs\,log\,g$
plane with evolution tracks for post-RGB stars. The consequent mass for
the secondary component was $0.066\pm\,0.001\,M_{\odot}$ which places
it firmly in the brown dwarf category.

With the discrepancy about the surface gravity still unresolved, we
decided to secure new complete light curves of AA Dor in the $V$ and
$I$ bands in order to re-determine the radii of both stars relative to
the semi-major axis of the relative orbit. Most recently, Rauch \&
Werner (2003) have published a new spectroscopic study of AA~Dor based
upon many 180s-exposure UVES spectrograms obtained with the ESO Very
Large Telescope.

\section{Photometric Observations}

Because the primary eclipse of AA~Dor has a total duration of only 26
minutes and a depth of 0.4 mag, it is clear that photometry has to be
secured with excellent time resolution in order to ensure that the
detailed shape of the eclipse curve is established without
compromise. Accordingly, the University of Capetown CCD camera,
mounted on the 1.0-m telescope at the South African Astronomical
Observatory, was used to secure these data. This camera operates in
frame-transfer mode, so that there is essentially no time lost between
exposures. Integration times of 15- or 20-s were used for the $V$
observations, and 20-s for the $I_{C}$ observations, ensuring that the
eclipse curves were well sampled.  To be certain of this time
resolution, all the observations on a single night were obtained in
one filter only, rather than alternating between the $V$ and $I_{C}$
passbands.  The field of view of the UCT CCD camera on the 1-m telescope
is only about $1.8\times\,1.2$ arcminutes so that,
despite AA~Dor being a foreground object to the Large Magellanic
Cloud, the field is remarkably free of useful comparison stars. One comparison
star is $\simeq\,1m$ fainter than AA~Dor at maximum, and the rest are all at
least $2m$ fainter.

Observations of the AA~Dor field were obtained on 4 consecutive
nights, two nights with incomplete light curves, and two nights with
complete light curves, one in each of the $V$ and $I_{C}$ passbands
that are in routine use at the SAAO. The data were reduced with an
SAAO version of DoPhot, slightly modified to reduce data ``on-line''
so that an almost immediate assessment of photometric quality can be
made.  Differential magnitudes were derived relative to the brightest
comparison star using standard SAAO software. The constancy of this
star was established by comparing it with the fainter stars in the
field.

Infrared $JHK$ photometry was obtained using the MkIII photometer
(Glass 1973), with the $f/50$ chopping secondary, on the 1.88-m
reflector of the SAAO. Observations were made during two allocations of
time: 1984 January 17--23 and 1984 January 27--30. For each filter,
observations were obtained in modules of two 10-s integrations.
Beamswitching occurred twice in each module, which was repeated as
necessary in order to achieve the desired accuracy. On a given night,
a single filter was selected and AA~Dor monitored for an entire orbital
period. The star HD37027 was selected as a comparison star since it was
an early-type star close to AA~Dor and which Cousins (1987) subsequently
published as a secondary $UBV$ standard star.

Infrared standard stars were selected from the SAAO list, which 
Carter (1993,1995) later published, and observed in so far as was practicable
at approximately the same airmass as AA~Dor. These enabled differentially-
corrected infrared magnitudes to be placed on the standard system. The
intention was to detect a signature of the cool secondary component
in the infrared light curves, specifically by examining whether the depth
of the secondary eclipse would be found to dip below the occultation
of the reflection effect by the primary, that is, below the level
recorded immediately outside the primary eclipse, and thereby yield
some measurement of the infrared flux from the secondary star. The data
were of poorer quality than had been hoped, and they failed to show a clear
signature of the flux from the secondary star beyond the known reflection
effect; they are included graphically as a further test of reflection-effect
modelling, considered in section 3.3.

\section{Analysis of the light curves}
\subsection{The $VI_{C}$ data}

The most recent orbital ephemeris for AA~Dor has been determined by
Kilkenny \textit{et al.} (2000) from times of primary eclipse minima
observed from 1977 to 1999. This linear ephemeris has a reference time
of primary minimum of $HJD\,2443196.34868\pm\,0.00002$, and an orbital
period of $0.261539731\pm\,0.000000002\,d$, and has been used to
convert the heliocentric Julian Dates into orbital phases. To improve
photometric accuracy, the individual observations were added in
orbital-phase bins of $\Delta\,\phi\,=0.0020$ and then differential
magnitudes formed between the variable and the comparison star. Most
bins had 2-3 observations yielding a typical Poisson uncertainty of
$\pm0.004\,m$ in $V$ and $\pm0.003\,m$ in $I_{C}$.

\begin{figure}
 \includegraphics[width=84mm]{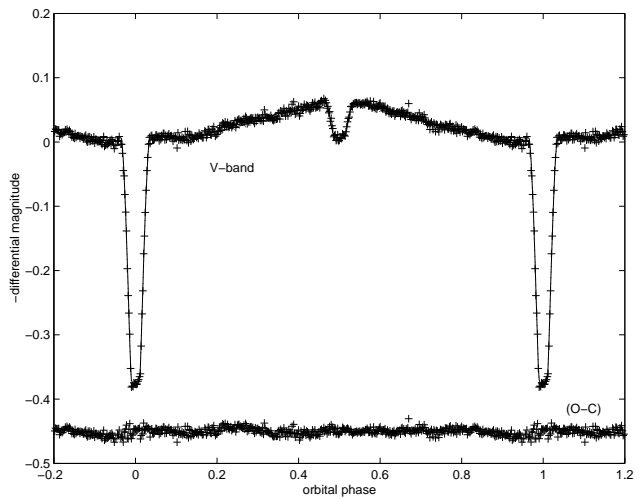} 
\caption{The complete $V$ band
 light curve of AA~Dor defined by 500 binned observations (crosses
 with Poisson uncertainties of $\pm0.004\,m$). The solid line
shows the  \textsc{light2} model fit to these data, whilst the dotted line
is the theoretical curve with the albedo fixed at unity.
The observed-calculated $(O-C)$  values are also plotted displaced
by $+0.45\,m$ from zero.}
 \label{Vlc}
\end{figure}

\begin{figure}
 \includegraphics[width=84mm]{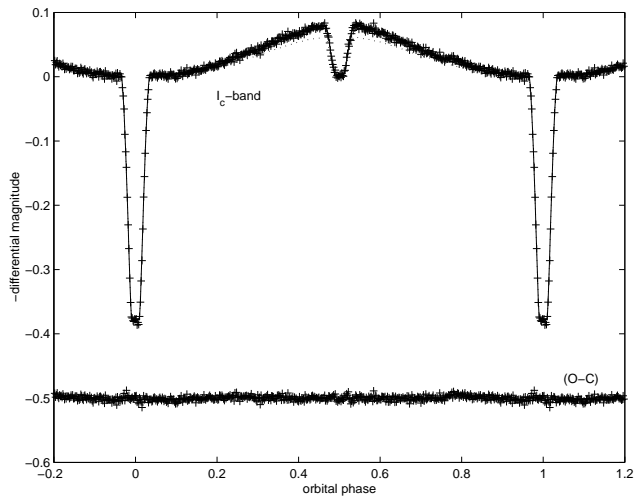}
 \caption{The complete $I_{C}$ band light curve of AA~Dor defined by
500 binned observations (crosses with Poisson uncertainties of 
$\pm0.003\,mag$). The solid line shows the \textsc{light2} model
fit to these data, whilst the dotted line is the theoretical curve with
the albedo fixed at unity.
The observed-calculated $(O-C)$  values are also plotted displaced
by $+0.50\,m$ from zero. }
 \label{Ilc}
\end{figure}

The two complete light curves are plotted in Figure~\ref{Vlc}, with a
total of 500 binned observations defining the $V$ curve from 1157
individual observations obtained on JD2451880, and in Figure~\ref{Ilc}
with a total of 500 binned observations defining the $I_{C}$ curve
from 1287 individual observations obtained on JD2451879.  On each of
these nights complete light curves were secured and the small region
of orbital-phase overlap with data obtained at the beginning and end
of each night showed no differences. The overall $V$ curve shows some
minor irregularities from a symmetric curve, but the $I_{C}$ curve is
entirely symmetric.

\subsection{Geometric parameters}

The two stellar radii, expressed in terms of the semi-major axis of
the relative orbit, and the orbital inclination are determined very
well from the annular primary eclipse curve alone, as noted in HHH.
We have used the \textsc{light2} light-curve-synthesis code 
(Hill 1979, Hill \& Rucinski 1993) to analyse
the complete $V$ and $I_{C}$ curves separately, and the results are
given in Table~\ref{geompar}. We adopted standard values of linear
limb darkening for both stars, together with standard values of the
gravity darkening exponent. A mass ratio of 0.20 (secondary/primary)
was adopted from Rauch (2000).  The effective temperature of the sdO
star was taken to be $42000K$ also from Rauch's (2000) NLTE spectral
analysis, and that of the cool secondary star to be a nominal
$2000K$. Solutions of the $I_{C}$-band curve with different fixed
values of the secondary temperature ($T_{sec}$) showed that the
$\chi^{2}$ value increases by a factor 2 over the range
$2000K<T_{sec}<6000K$. The reason for this lack of dependence on the
secondary temperature is simply that the flux ratio
(primary/secondary) is $10^{4}$ in the $I_{C}$ band and $10^{5}$ in
the $V$ band immediately after egress from the primary eclipse, when
we see only the averted unheated hemisphere of the secondary. At
the maximum of the reflection effect this ratio has reduced to 13 in
$I_{C}$ and 19 in $V$, which suggests that it should be possible to
detect the heated face of the secondary in near-infrared spectra
obtained at the appropriate orbital phases.

\begin{table}
\caption{Geometric parameters}
\label{geompar}
\begin{tabular}{crrrr}
$passband$&$\overline{r_{pri}}$&$\overline{r_{sec}}$&$incl$&$rms$ \\
\hline
$V$ &$0.1413$&$0.0759$&$89.60$&$\pm\,0.0053$ \\
    &$\pm0.0008$&$\pm0.0006$&$\pm0.56$&$ $ \\
$I_{C}$&$0.1425$&$0.0769$&$88.82$&$\pm\,0.0039$ \\
     &$\pm0.0007$&$\pm0.0005$&$\pm0.20$&$ $ \\
\hline
$average$&$0.1419$&$0.0764$&$89.21$&$ $ \\
    &$\pm0.0005$&$\pm0.0003$&$\pm0.30$&$ $ \\  
\hline 
\end{tabular}
\end{table}

It is clear from the results in Table ~\ref{geompar} that the
solutions for the mean, or volume, radii and the inclination from the
$V$ and $I_{C}$ curves are in excellent agreement, and that the $I_{C}$
solution is slightly more precise. The average values have been
adopted in all subsequent analyses as fixed parameters, and we should
note that they are little different from those found in
earlier studies from less complete and less accurate data, but they
are more precisely determined.

\begin{figure}
 \includegraphics[width=84mm]{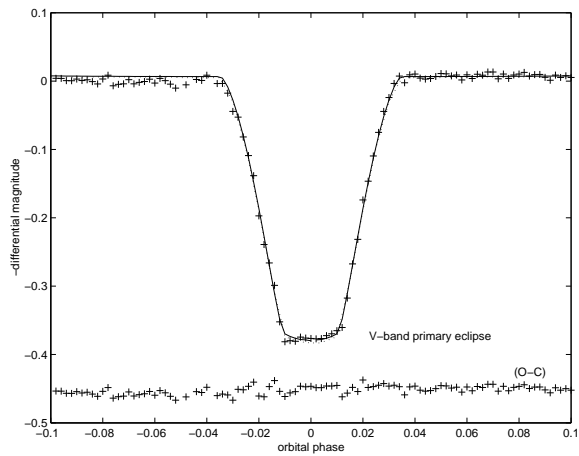}
 \caption{The $V$ band primary eclipse light curve of AA~Dor 
(crosses with Poisson uncertainties of $\pm0.004\,m$). 
The solid line is the \textsc{light2} model
fit to these data solving for the geometrical parameters and the albedo
of the secondary star. The dotted line is a theoretical curve calculated from
the average geometrical parameters and adopting an albedo of unity for the
secondary.The observed-calculated $(O-C)$  values are also plotted displaced
by $+0.45\,m$ from zero. }
 \label{Vprecl}
\end{figure}

\begin{figure}
 \includegraphics[width=84mm]{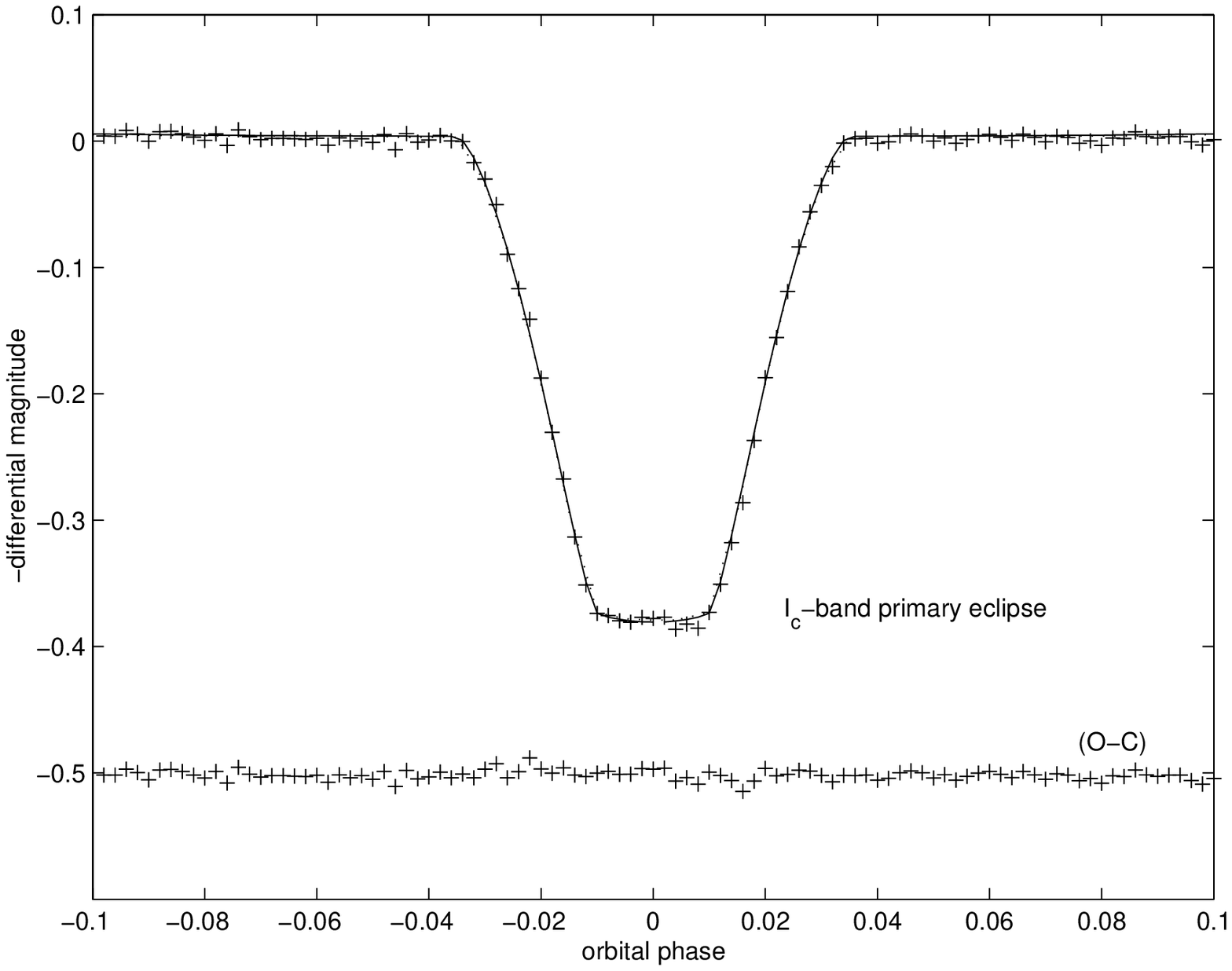} 
\caption{The $I_{C}$ band
 primary eclipse light curve of AA~Dor (crosses with Poisson
 uncertainties of $\pm0.003\,m$).  The solid line is the
 \textsc{light2} model fit to these data solving for the geometrical
 parameters and the albedo of the secondary star. The dotted line is a
 theoretical curve calculated from the average geometrical parameters
 and adopting an albedo of unity for the secondary.The
 observed-calculated $(O-C)$ values are also plotted displaced by
 $+0.50\,m$ from zero. } 
\label{Iprecl}
\end{figure}

\begin{figure}
 \includegraphics[width=84mm]{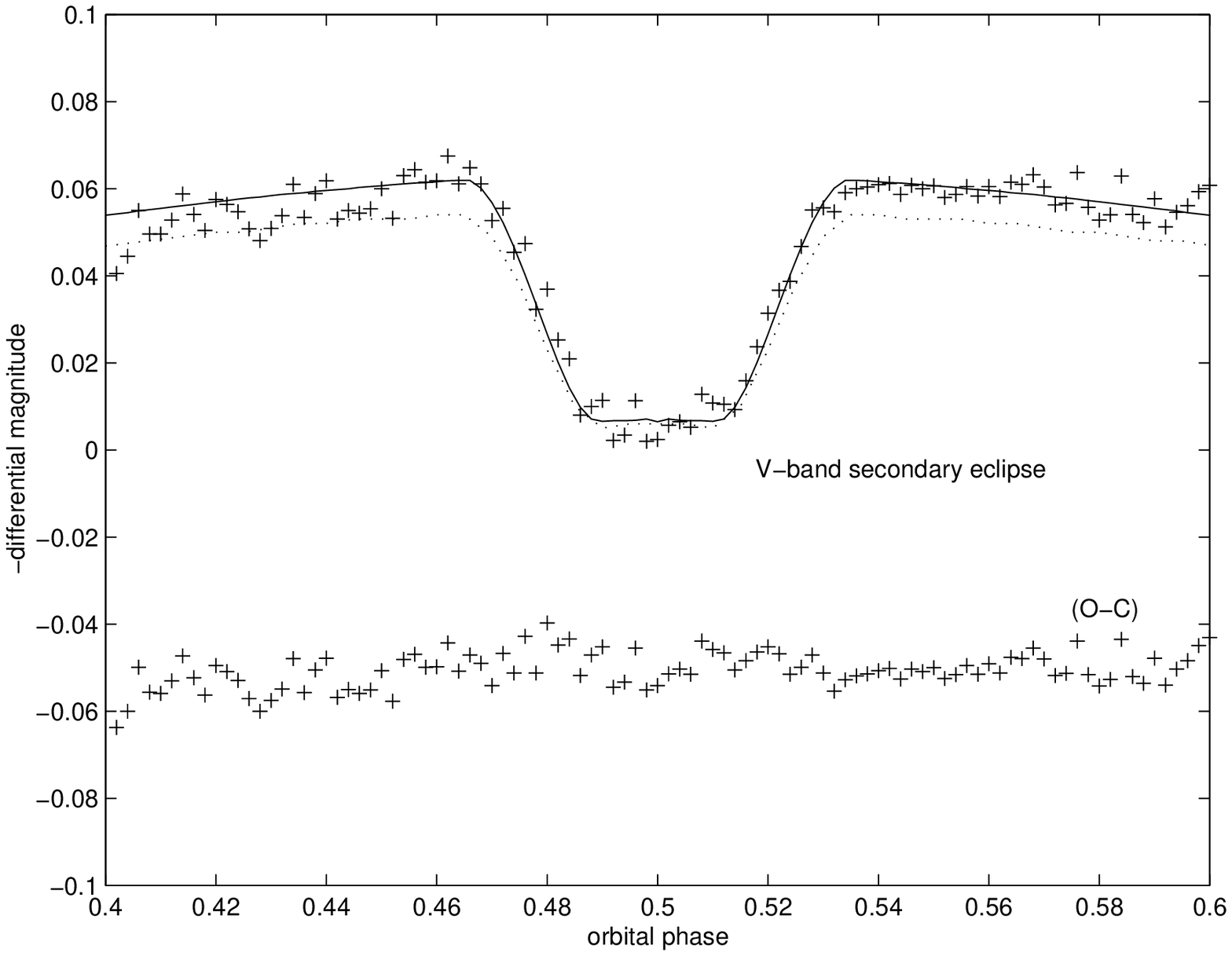}
 \caption{The $V$ band secondary eclipse light curve of AA~Dor
 (crosses with Poisson uncertainties of $\pm0.004\,m$). 
The solid line is the \textsc{light2} model
fit to these data solving for the geometrical parameters and the albedo
of the secondary star. The dotted line is a theoretical curve calculated
from the average geometrical parameters and adopting an albedo of unity for
the secondary.The observed-calculated $(O-C)$  values are also plotted 
displaced by $+0.050\,m$ from zero. }
 \label{Vsececl}
\end{figure}

\begin{figure}
 \includegraphics[width=84mm]{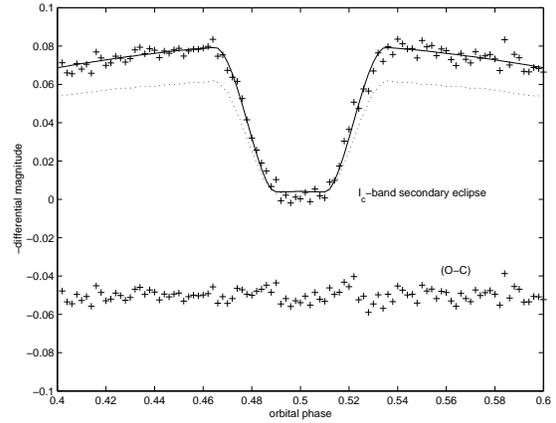}
 \caption{The $I_{C}$ band secondary eclipse light curve of AA~Dor
 (crosses with Poisson uncertainties of $\pm0.003\,m$). 
The solid line is the \textsc{light2} model
fit to these data solving for the geometrical parameters and the albedo
of the secondary star. The dotted line is a theoretical curve calculated
from the average geometrical parameters and adopting an albedo of unity
for the secondary.The observed-calculated $(O-C)$ values are also plotted 
displaced by $+0.050\,m$ from zero. }
 \label{Isececl}
\end{figure}

The excellent agreement between these $V$ and $I_{C}$-band
observations and the model light curves are shown in detail through
the primary and secondary eclipse curves in Figure~\ref{Vprecl},
Figure~\ref{Iprecl}, Figure~\ref{Vsececl} and Figure
~\ref{Isececl}. $V$ and $I_{C}$ curves defined by the individual
observations were also solved, and the same values for the geometrical
parameters were determined as in Table 1 but with larger
uncertainties. Combining the observations into these phase-binned
values has not systematically affected the derivation of the relative
stellar radii.

\subsection{Reflection effect}

The manner in which the reflection effect is modelled does not
significantly affect the solution for the geometrical parameters in
AA~Dor, which is why that part of the analysis has been separated.
In HHH, it was explained that the \textsc{light2} code had difficulty
in fitting the observed reflection effect, and recourse was made to
rather extreme measures to fit the data, by changing the limb-darkening
coefficient of the heated secondary star to negative (limb-brightening)
values. Whilst such drastic action did provide a solution, it was clearly
not satisfactory. 

With these new complete light curves, it was necessary to resolve this
issue, particularly in view of the fact that users of the
Wilson-Devinney (\textsc{wd}) code (Wilson \& Devinney 1971, 
Wilson 1979, 1990) on eclipsing binaries with similar
large-amplitude reflection effects did not experience such major
difficulties (e.g. Wood \textit{et al.} 1993, Zola 2000), although it did
seem necessary for Wood \textit{et al.} to adopt some unusually low values
for the limb-darkening coefficients.  A detailed reexamination of the
code by GH finally revealed a one-line error in the reflection
calculation subroutine, a missing factor $\pi$ coupled with an incorrect
expression involving the angle to the surface normal of emergent intensity.
Making this correction increased the calculated amplitude of the 
reflection effect up to that calculated by the \textsc{wd} code.

This revised code was tested against a set of light curves kindly
provided by Professor R.E.Wilson (private communication). These
monochromatic light curves were generated with the current version of
the \textsc{wd} code using nominal values of the two radii, the
orbital inclination, and the temperatures of the two stars in AA~Dor,
and setting the bolometric albedo of the secondary star to be unity.
The monochromatic wavelengths were set to the mean wavelengths of the
Str\"{o}mgren-Crawford $uvby$ filters, the $V$ and $I_{C}$ filters,
and the infrared $JHK$ passbands.  Since it is already
long-established that the \textsc{wd} and \textsc{light2} codes agree
in all geometrical respects, (cf., for example, Bell \textit{et al.}
1991), these $uvby$ and $VI_{C}$ light curves were solved individually
by the revised \textsc{light2} code for the albedo of the secondary
component. Two descriptions of the reflection effect were used, one
adopting that in the \textsc{wd} code of complete absorption and
re-radiation of the incoming flux from the primary star by the
secondary star (i.e. heating), and the other that provided by
Sobieski(1965) where some of the incoming flux may be scattered by
free electrons in the atmosphere of the secondary and the remainder
goes into direct heating. The direct heating approach yielded a mean
albedo from the 9 light curves of $0.97\pm0.06$, whilst the Sobieski
approach yielded a mean albedo of $1.16\pm0.07$. This direct
comparison between the \textsc{wd} and the \textsc{light2} code shows
a satisfactory agreement in the representation of the reflection
effect.

The $V$ and $I_{C}$ light curves were analysed with the \textsc{light2}
code, adopting black-body fluxes convolved with the standard filter
response curves, and solving for the two radii, the orbital
inclination, and the albedo ($\alpha_{sec}$) of the secondary
component. Because of the large range in temperature across the heated
hemisphere of the secondary component ($2,000-20,000\,K$), we used
the option in \textsc{light2} to use a limb-darkening
coefficient appropriate to the local temperature of each code pixel on the
secondary surface rather than a mean value.  
Both modes of calculating the reflection effect were used,
and these provided identical values of the geometrical
parameters. These values are reported in Table~\ref{geompar}, as
already noted. The values for $\alpha_{sec}$ are given in
Table~\ref{albedos} for the two wavelengths and both modes of
calculating the reflection effect. The root-mean-square (\textit{rms}) 
scatter of the binned observations about the fitted light curves 
(noted in Table~\ref{geompar}) are only $0.001\,m$ larger than the
Poisson uncertainties on these binned data, demonstrating that the fits
are very good.

\begin{figure}
 \includegraphics[width=84mm]{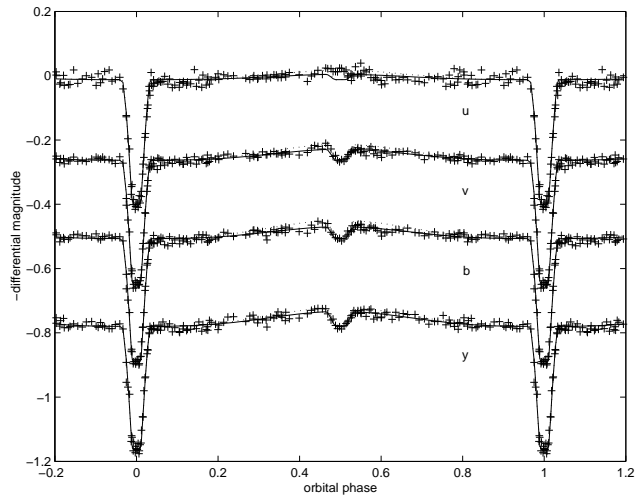}
 \caption{The $uvby$ light curves of AA~Dor (crosses) published by
Kilkenny {\em et al.}(1979), and the \textsc{light2} model fit to these data 
(solid lines) adopting the average geometrical parameters and solving
for the secondary albedo. The dotted lines show the theoretical curves
for an albedo of unity. }
 \label{uvbylcs}
\end{figure}

The $uvby$ light curves of AA~Dor published by Kilkenny, Penfold \&
Hilditch (1979) (Figure~\ref{uvbylcs}) were also analysed with the
\textsc{light2} code solving only for the albedo of the secondary
component in both the heating-only mode and in the Sobieski mode, and
adopting as fixed values the average geometrical parameters given in
Table 1 from the $V$ and $I_{C}$ light curve solutions.  These
results, together with those determined from the $V$ and $I_{C}$ curve
solutions are listed in Table~\ref{albedos}.

\begin{table}
\caption{Values of heating albedo for the secondary component}
\label{albedos}
\begin{tabular}{ccc}
$passband$&$heating-only$&$Sobieski$ \\
\hline
$ u $&$ 0.46 $&$ 0.49 $ \\
$   $&$\pm0.12$&$\pm0.16$ \\
$ v $&$ 0.45 $&$ 0.48 $ \\
$   $&$\pm0.06$&$\pm0.06$ \\
$ b $&$ 0.39 $&$ 0.42 $ \\
$   $&$\pm0.05$&$\pm0.06$ \\
$ y $&$ 0.55 $&$ 0.63 $ \\
$   $&$\pm0.07$&$\pm0.10$ \\
$ V $&$ 1.43 $&$ 1.70 $ \\
$   $&$\pm0.05$&$\pm0.06$ \\
$I_{C}$&$ 2.02 $&$ 2.40 $ \\
$     $&$\pm 0.04$&$\pm0.05$ \\
\hline 
\end{tabular}
\end{table}

Irrespective of the mode of reflection calculation used, there are
significant departures from an idealised $\alpha_{sec}=1.0$ model. In
Figures 3-7, these solution curves are plotted as solid lines, there
being no difference between the two modes. Also plotted in these
figures are theoretical light curves for a binary with the average
geometrical parameters and $\alpha_{sec}=1.0$. The $uvby$ light curves
give an average value of $\alpha_{sec}=0.50\pm0.05$ for the Sobieski
mode and $\alpha_{sec}=0.46\pm0.04$ for the heating-only mode. The
$\alpha_{sec}=1.0$ light curves systematically do have reflection
effects with larger amplitudes than the observations, but those
differences amount only to $\simeq\,0.01-0.02\,m$, and the
observational uncertainties are of the same order per observation. For
the better-defined $V$ and $I_{C}$ curves, the observations present a
reflection effect of higher amplitude than the $\alpha_{sec}=1.0$
curves and these systematic differences are clearly evident,
particularly in Figure~\ref{Vsececl} and Figure~\ref{Isececl}. Since
these single epoch $V$ and $I_{C}$ light curves were obtained 22 years
after the $uvby$ data, we cannot rule out the possibility of some
intrinsic variability in the system that conspires to simulate an
enhanced reflection effect in two wavelengths and a suppressed effect
at shorter wavelengths. Alternatively, perhaps this is observational
evidence for some redistribution of flux from the near-UV and blue
regions into the visual and near-IR regions by means of substantial
line blocking/blanketing. Note that this intrinsically very cool
secondary star has an unperturbed $T_{eff}=2000\,K$, and a temperature
at the substellar point directly underneath the sdOB primary star of
some $20,000\,K$. It would be interesting to know what type of overall
spectrum would be calculated for such an illuminated hemisphere.
($cf.,e.g.$,Barman, Hauschildt \& Allard, 2002). In our analysis, the
black-body fluxes are a reasonable representation of the broad-band
sdOB spectrum. But the spectrum of the heated hemisphere of the
secondary component is likely dominated by emission lines and our
black-body representation is clearly a limitation. It might also be
expected that circulation currents would exist on the facing
hemisphere of the irradiated secondary star (Podsiadlowski 1991), and
that the limb-darkening values would be reduced compared to normal.
Reducing the limb-darkening values increases the calculated amplitude
of the reflection effect for a given value of the albedo, as already
demonstrated by Wood {\em et al.}, and by HHH.

\begin{figure}
 \includegraphics[width=84mm]{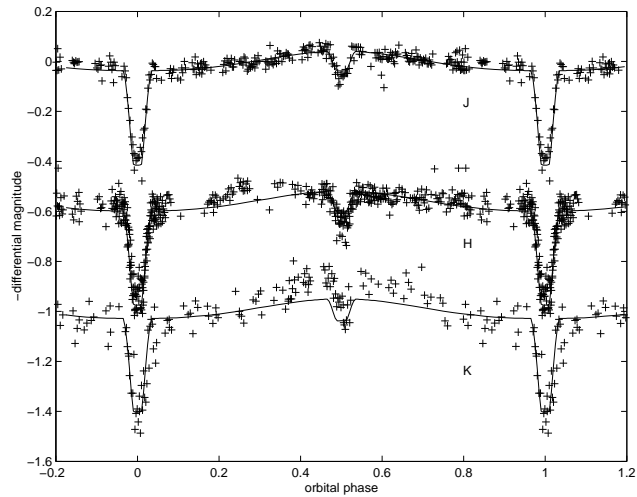} 
\caption{The $JHK$ light curves of AA~Dor (crosses) obtained by AELG. The
solid lines show the theoretical curves calculated from the average 
geometrical parameters for an albedo of unity.}  
\label{jhklcs}
\end{figure}

In Figure~\ref{jhklcs}, the infrared JHK band data are plotted, together
with theoretical curves calculated from the average geometrical parameters
and with $\alpha_{sec}=1.0$. Despite the obvious large uncertainties in
these data, such theoretical light curves are consistent with these data
and do not suggest any large-scale departure from a model with a 
bolometric albedo of unity for the secondary star.

We note again that Wood \textit{et al.} experienced some difficulties
in fitting their multi-colour observations of HW~Vir with the 
\textsc{wd} code in the sense that they had to adopt low values of the
limb-darkening coefficients in order to increase the theoretical
amplitude of the reflection effect up to that observed. In a study of the
system PG1336-018, Zola (2000) modified the \textsc{wd} code to assign
limb-darkening coefficients appropriate to the local temperatures of 
each of the stellar surface pixels in the code, and obtained an albedo
value in the range $0.45-0.95$ dependent on other parameters for the 
system. Thus there is evidence that in these binary systems displaying
extreme reflection effects the accepted standard light-curve-synthesis
codes experience some difficulties in matching the observations.  

\section{Astrophysical parameters}

The results of several observational programmes investigating the
fraction of hot subdwarf stars that are found in binary systems have
been published in recent years. Specifically, Maxted, Heber, Marsh \&
North (2001) and many references therein, have demonstrated that the
observed fraction of extreme horizontal branch stars that are members
of binary systems with orbital periods less than 10 days is $\simeq\,70\%$.
Two recent papers by Han {\em et al.} (2002) and Han, Podsiadlowski,
Maxted \& Marsh (2003) have discussed the various evolution channels
that can explain the observed properties of sdB,sdOB, and sdO stars in
binary systems, concluding that the different channels can explain why
some hot subdwarfs are found in short-period binaries, others in
binaries with much larger orbits, and some are now merged single
objects. The first common-envelope-ejection (CEE) channel is the evolution
path where the initially more massive star experiences dynamical mass
transfer on its first ascent of the red-giant branch, leading to the
formation of a common-envelope followed by a spiral-in phase with the
binary system attaining a short orbital period ($P<10d$) and being
composed of a hot subdwarf and a main-sequence companion. The
distribution of masses of the hot subdwarf stars in this evolution
channel is expected to lie in the range $0.33-0.47\,M_{\odot}$, with
a main peak at around $0.46\,M_{\odot}$ and a much smaller peak around
$0.33\,M_{\odot}$. These binary population synthesis models do succeed
in explaining the observed distributions of properties of sdB,sdOB,sdO
star binaries. Specifically with regard to AA~Dor, the first CEE models
seem to provide a sensible explanation for its observed properties.

Since the spectrum of the secondary star in AA~Dor has not yet been 
observed, so that we do not know its orbital velocity curve, we have to
assume a mass for the sdOB star, and then calculate all the absolute
astrophysical parameters from the derived values of relative radii, orbital
inclination, and the velocity semi-amplitude of the primary component.
HHH derived a value of $K_{pri}=40.8\pm0.7\,km\,s^{-1}$ whilst Rauch \&
Werner (2003) determined a value of $K_{pri}=39.19\pm\,0.05\,km\,s^{-1}$.
Adopting an expected range of subdwarf mass of $0.33-0.47\,M_{\odot}$,
with both values of $K_{pri}$, and the geometric parameters from 
Table~\ref{geompar}, we obtain the ranges of possible values of
masses, radii, surface gravities, and synchronous rotational velocities
given in Table~\ref{abspar}.

\begin{table}
\caption{Astrophysical parameters}
\label{abspar}
\begin{tabular}{ccc}
$parameter$&$primary$&$secondary$ \\
\hline
$mass\,M_{\odot}$&$0.33-0.47$&$0.064-0.082$ \\
$radius\,R_{\odot}$&$0.179-0.200$&$0.097-0.108$ \\
$log~g\,(cgs)$&$5.45-5.51$&$5.27-5.29$ \\
$V_{rot}\,km\,s^{-1}$&$34.7-38.6$&$18.7-20.8$ \\
\hline 
\end{tabular}
\end{table}

The corresponding range of mass ratio is $0.168-0.200$ and has no
effect upon the derived relative radii, orbital inclination etc from
the light curve analyses.  These ranges result from the assumed range
of primary star mass, and are larger than the range of uncertainties
that would result from propagation of the formal uncertainties of the
derived quantities. They may be compared with the values determined by
Rauch (2000) and Rauch \& Werner (2003) from their analyses of spectra
of AA~Dor.

Rauch (2000) derived a mass for the sdOB star of
$0.330\pm\,0.006\,M_{\odot}$ from comparison of their derived
effective temperature ($T_{eff}=42000\pm\,1000\,K$), and surface
gravity ($log~g=5.21\pm\,0.10$) with theoretical models for the
evolution of subdwarf stars of different masses plotted in the
$T_{eff}\,vs\,log\,g$ plane. This value of surface gravity disagrees
with that determined from the velocity curve of the primary component
and the above light curve analysis. We note that Rauch \& Werner
(2003) state:``Since the analysis of Rauch (2000) was hampered by the
relatively long exposure times (1h and 2-3h, respectively) and hence,
a relatively large orbital velocity coverage (the observed line
profiles were broadened by the star's rotation as well as by smearing
due to orbital motion within the observations), it has been speculated
that $log\,g$ from Rauch (2000) is somewhat too low.'' The new data
reported in Rauch \& Werner (2003) (hereafter RW) have exposure times
of just 180s and remove this problem. From NLTE model atmospheres, RW
fitted theoretical line profiles for the $HeII\lambda\,4686$
absorption line and derived $T_{eff}=44000\,K$, $log\,g=5.4$, without
published uncertainties, and $V_{rot}=43\pm\,5\,km\,s^{-1}$.  They
note, however, that this determination of $T_{eff}$ is not as
definitive as that determined by Rauch (2000) from the ionization
equilibria of many species, whilst the value of $log\,g=5.21 $ was
determined from the broad Balmer lines of hydrogen. Later, RW note
that: `` Unfortunately, significant effects of changes in $g$ are
detectable only in the outer line wings where the data reduction of
the broad Balmer lines in the echelle spectra is not very accurate.''
The value of $log\,g=5.4$ for $HeII\lambda\,4686$ is remarkably close
to the range in Table~\ref{abspar} derived from our radial-velocity
and light-curve analyses, and the value of
$V_{rot}=43\pm\,5\,km\,s^{-1}$ is consistent with synchronism.

These new results from RW suggest that the apparent discrepancy in
surface gravity may already be resolved (cf. the above results from
the $HeII\lambda\,4686$ line), although RW did reject this solution in
favour of the Rauch (2000) value. With $log\,g=5.21$  and $T_{eff}=42000K$
fixed, RW determined a rotational velocity of $V_{rot}=47\pm\,5\,km\,s^{-1}$
from the $\lambda\,4686$ line, which is $~20-35\%$ faster than synchronism.

The ranges of mass, radius and surface gravity for the sdOB star lie
within the expectations of the binary population synthesis models of
Han {\em et al.}, whilst the ranges of mass, radius and surface
gravity of the unseen secondary component are consistent with models
for the lowest mass dwarfs (Dorman {\em et al.} 1989) at the boundary
of the lower main sequence and the substellar brown-dwarf region.

The referee has commented that the $T_{eff},log\,g$ values for the
primary component of AA~Dor place it in an unusual and
relatively-short-lived phase of evolution compared to most hot
subdwarf stars. When compared to the binary-star evolution models of
Han {\em et al.} (2002), or the single-star evolution models of Dorman, Rood
\& O'Connell (1993), its location would suggest that the AA~Dor
primary is evolved beyond the terminal-age extreme horizontal branch
(\textsc{taehb}) towards the white dwarf region, a phase of evolution that
lasts $\simeq\,10^{7}$ yr. We may ask whether the relative numbers of
sdO and sdOB stars to all hot subdwarf stars would be consistent with
that view, at least in the sense of being a small proportion of the
total.

From the analysis of Zone 1 of the Edinburgh-Cape (EC) Blue Object
Survey, Kilkenny et al (1997) found 27 sdOB stars (like the AA~Dor
primary), and 20 sdO stars out of a total of 357 identified hot
subdwarfs, making proportions of $\simeq\,8\%$ and $\simeq\,6\%$
respectively. The vast majority ($\simeq\,76\%$) are sdB stars, with
the next most common being the He-sdO and He-sdB stars at
$\simeq\,11\%$. From the Hamburg Schmidt Survey (HSS), Lemke et al
(1997) selected 400 stars for more detailed spectroscopic
analysis. They identified 34 He-sdO stars and 13 normal sdO stars, the
latter being $\simeq\,3\%$ of the total. Their derived values of
$T_{eff}$ for these normal sdO stars lay in the wide range $32,000K -
100,000K$, whilst the values of $log\,g$ were in the range $5.0 - 7.0$
with the average at $5.6\pm0.6$, very similar to the AA~Dor
primary. These statistical results, and the model-atmosphere analyses,
are consistent with the evolutionary expectations that sdO, sdOB stars
do lie beyond the \textsc{taehb} in this short-lived phase of evolution.  By
contrast, the very recent spectral analyses by Edelmann {\em et al.} (2003)
of sdB stars discovered in the HSS is in agreement with the work of
Maxted {\em et al.} (2001) on the locations of the sdB stars mostly within
the bounds between the zero-age and the terminal-age extreme
horizontal branch.

\section{Summary}

Revised values of the mean radii of the two stars in AA~Dor have been
determined from these new $VI_{C}$ light curves which differ only a
little from those determined earlier, but they are defined more
precisely.  These results together with the published orbital
parameters from HHH and RW provide new determinations of the
astrophysical parameters with an adopted range of primary star mass.
These results are compatible with the theoretical evolution models for
binary systems containing hot subdwarf stars published by Han {\em et
al.}  (2002, 2003).

An error in the reflection effect subroutine within the
\textsc{light2} code has been corrected with the result that it agrees
with the accepted standard \textsc{wd} code. Analysis of the observed
reflection effect demonstrates significant departures within some
passbands from an idealised albedo of unity for the heated hemisphere
of the secondary star, but the overall bolometric value is most likely
close to unity. It is suggested that further theoretical work should
be done on modelling the reflection effect in these types of binaries
containing stars with very different effective temperatures.

\section{Acknowledgments}

The authors are indebted to Professor I.D.Howarth who participated in
obtaining the infrared photometry, and the referee, Dr Ph. Podsiadlowski, for his
helpful comments on the first draft of this paper . This research made use of the 
\textsc{simbad} database, and funds from a PPARC Visiting Fellowship
grant for GH to visit the University of St Andrews, during which some of 
this work was completed.

\end{document}